\documentclass[onecoloumn]{aa}
\usepackage{graphicx}

\def\la{\;\raise0.3ex\hbox{$<$\kern-0.75em\raise-1.1ex\hbox{$\sim$}}\;}
\def\ga{\;\raise0.3ex\hbox{$>$\kern-0.75em\raise-1.1ex\hbox{$\sim$}}\;}

\begin{document}
   \title{
A new constraint on the time dependence of \\
the proton$-$to$-$electron mass ratio}

   \subtitle{Analysis of the Q~0347$-$383 and Q~0405$-$443 spectra}

   \author{A.~Ivanchik\inst{1} \and P.~Petitjean \inst{2,3}
            \and D.~Varshalovich\inst{1}
            \and B.~Aracil\inst{2,4}
            \and R.~Srianand \inst{5}
            \and H.~Chand \inst{5}
            \and C.~Ledoux \inst{6}
            \and P.~Boiss\'e \inst{2,3}
          }

   \offprints{A.~Ivanchik}

   \institute{Ioffe Physical Technical Institute, Polytekhnicheskaya 26, 194021 Saint-Petersburg, Russia \\
             \email {iav@astro.ioffe.rssi.ru}
\and
Institut d'Astrophysique de Paris -- CNRS, 98-bis Boulevard Arago F-75014, Paris, France  \\
             \email {ppetitje@iap.fr}
             \and
             LERMA, Observatoire de Paris, 61 avenue de l'Observatoire, F-75014, Paris, France
             \and
Department of Astronomy, University of Massachusetts,
710 North Pleasant Street, Amherst, MA 01003-9305, USA
             \and
             IUCAA, Post Bag 4, Ganeshkhind, Pune 411007, India
             \and
              European Southern Observatory, Alonso de C\'ordova 3107, Casilla 19001, Vitacura, Santiago, Chile \\
             }

   \date{Received xxxx x, 200x; accepted xxxx x, 200x}

   \abstract{
   A new limit on the possible cosmological variation of the proton-to-electron mass
   ratio $\mu=m_p/m_e$ is estimated by measuring wavelengths of H$_2$
   lines of Lyman and Werner bands from two absorption systems at $z_{\rm abs}=2.5947$ and
   $3.0249$ in the spectra of quasars Q~0405$-$443 and Q~0347$-$383, respectively.
   Data are of the highest spectral resolution ($R$~=~53000) and S/N ratio (30$\div$70) for this
   kind of study.
   We search for any correlation between  $z_{\rm i}$, the redshift of observed lines, determined
   using laboratory wavelengths as references, and $K_{\rm i}$,
   the sensitivity coefficient of the lines to a change of $\mu$, that could be interpreted
   as a variation of $\mu$ over the corresponding cosmological time.
   We use two sets of laboratory wavelengths, the first one, Set~(A) (Abgrall et al.~\cite{Abgr}),
   based on experimental determination of energy levels and the second one, Set~(P) (Philip et
   al.~\cite{Philip}), based on new laboratory measurements of some individual rest-wavelengths.
   We find $\Delta\mu$/$\mu$~=~(3.05$\pm$0.75)$\times$10$^{-5}$ for Set~(A),
   and  $\Delta\mu$/$\mu$~=~(1.65$\pm$0.74)$\times$10$^{-5}$ for Set~(P). The second determination
   is the most stringent limit on the variation of $\mu$ over the last 12~Gyrs ever obtained.
   The correlation found using Set~(A) seems to show that some amount of systematic error is
   hidden in the determination of energy levels of the H$_2$ molecule.

   }

   \maketitle
%

\section{Introduction}

Contemporary theories of fundamental interactions
(Strings/M-theory and others) predict some variation of the
fundamental physical constants in the course of the evolution of the Universe.
Most of the predictions of such theories lie in the
energy range inaccessible to current experiments ($E \sim10^{19}$
GeV). However, at lower energy, variations of the fundamental
constants, in principle, could be a possible observational
manifestations of these theories. It is therefore important to
constrain these variations as a step toward a better understanding
of Nature.

A considerable amount of interest in the possibility of time
variations of fundamental constants has been generated by recent
observations of quasar absorption systems. Using a new method, the
so-called Many-Multiplet analysis (Webb et al., \cite{Webb1};
Dzuba et al., \cite{Dzuba}), Murphy et al. (\cite{Murphy03}) have
claimed that the fine structure constant, $\alpha=e^2/\hbar c$,
could have varied over the redshift range $0.2<z<3.7$ with the
amplitude $\Delta\alpha/\alpha=(-0.543\pm0.116)\times10^{-5}$.
However, a stringent upper limit on the variation of $\alpha$ has
been obtained from a large sample of UVES data,
$\Delta\alpha/\alpha=(-0.06\pm0.06)\times10^{-5}$
over $0.4<z<2.3$ (Srianand et al. 2004, Chand et al. 2004).
In addition, Quast et al. (\cite{Quast}) derived
$\Delta\alpha/\alpha=(-0.04\pm0.19\pm0.27)\times10^{-5}$ from an
analysis of one system at $z_{\rm abs}=1.15$.

One way to solve the controversy is to constrain other fundamental
constants. Different theoretical models of the fundamental
physical interactions predict different variations of their values
and different relations between cosmological deviations of the
constants ($\alpha$, $\mu$, and others, see Calmet \& Fritzsch
\cite{Calmet}, Langacker et al., \cite{Langacker}, Olive et al.
\cite{Olive}, Dent \& Fairbairn \cite{Dent}). Therefore, it is
crucial to couple measurements of different dimensionless
fundamental constants.

\section{The proton-to-electron mass ratio $\mu=m_p/m_e$}

Here we use QSO absorption lines to constrain $\,\Delta \mu/\mu$
with $\,\Delta \mu = \mu - \mu_0$, where $\,\mu\,$ is the
proton-to-electron mass ratio at the epoch of the QSO absorption
spectrum formation and $\mu_0$ is its contemporary value.

In the framework of unified theories (e.g. SUSY GUT) with a common
origin of the gauge fields, variations of the gauge coupling
$\alpha_{GUT}$ at the unified scale ($\sim10^{16}$~GeV) will
induce variations of all the gauge couplings in the low energy
limit, $\alpha_i=f_i(\alpha_{GUT},E)$, and provide a relation
$\Delta\mu/\mu\simeq R \Delta\alpha/\alpha$, where $R$ is a model
dependent parameter and $|R|\la 50$ (e.g. Dine et al.,
\cite{Dine} and references therein). Thus, independent estimates of $\Delta
\alpha/\alpha$ and $\Delta \mu / \mu$ could constrain the mass
formation mechanisms in the context of unified theories.

At present the proton-to-electron mass ratio has been measured
with a relative accuracy of $2\times 10^{-9}$ and equals $\mu_0 =
1836.15267261(85)$ (Mohr \& Taylor, \cite{Mohr}). Laboratory
metrological measurements rule out considerable variation of $\mu$
on a short time scale but do not exclude its changes over the
cosmological scale, $\sim 10^{10}$ years. Moreover, one can not
reject the possibility that $\mu$ (as well as other constants)
could be different in widely separated regions of the Universe.

The method used here to constrain the possible variations of $\mu$
was proposed by Varshalovich and Levshakov (\cite{VarshLev}). It
is based on the fact that wavelengths of electron-vibro-rotational
lines depend on the reduced mass of the molecule, with the
dependence being different for different transitions. It enables
us to distinguish the cosmological redshift of a line from the
shift caused by a possible variation of $\mu$.

Thus, the measured wavelength $\lambda_i$ of a line formed in the
absorption system at the redshift $z_{abs}$ can be written as
\begin{equation}
\lambda_{\rm i}=\lambda_{\rm i}^0(1+z_{abs})(1+K_{\rm i} \Delta\mu/\mu)
\label{era}
\end{equation}
where $\lambda_{\rm i}^0$ is the laboratory (vacuum) wavelength of the
transition, and $K_{\rm i}= \mbox{d} \ln \lambda_{\rm i}^0/ \mbox{d} \ln \mu$
is the sensitivity coefficient calculated for the Lyman and Werner
bands of molecular hydrogen in work (Varshalovich and Potekhin,
\cite{VarshPo}). This expression can be represented in terms of
the individual line redshift $z_{\rm i} \equiv \lambda_{\rm i}/\lambda_{\rm i}^0-1$ as
\begin{equation}
z_{\rm i}=z_{abs}+b K_{\rm i}
\label{korr}
\end{equation}
where $b=(1+z_{abs})\Delta\mu/\mu$. Note, in case of nonzero
$\Delta\mu/\mu$, $z_{\rm abs}$ is not equal to the standard mean
value $\overline{z}= (\Sigma \, z_{\rm i})/N$.

In reality, $z_{\rm i}$ is measured with some uncertainty which is
caused by statistical errors of the astronomical measurements
$\lambda_{\rm i}$, by errors of the laboratory measurements of
$\lambda_{\rm i}^0$, and by possible systematic errors.
Nevertheless, if $\Delta \mu / \mu$ is nonzero, there must be a
correlation between $z_{\rm i}$ and $K_{\rm i}$ values. Thus, a
linear regression analysis of these quantities yields $z_{\rm
abs}$ and $b$ (as well as their statistical significance),
consequently an estimate of $\Delta\mu/\mu$.

Previous studies have already yielded tight upper limits on
$\mu$-variations, $|\Delta\mu/\mu|<7\times10^{-4}$ (Cowie \&
Songaila, \cite{Cowie}), $|\Delta\mu/\mu|<2\times10^{-4}$
(Potekhin et al., \cite{PIV}), $|\Delta\mu/\mu|<5.7\times10^{-5}$
(Levshakov et al., \cite{Lev1}), and
$\Delta\mu/\mu=(3.0\pm2.4)\times10^{-5}$ (Ivanchik et al.,
\cite{IAV}). Using new laboratory measurements of H$_2$
wavelengths (Philip et al., \cite{Philip}) and previous qso data,
Ubachs \& Reinhold (\cite{Ubachs}) found
$\Delta\mu/\mu=(-0.5\pm1.8)\times10^{-5}$.

\section{Observations}
\begin{figure*}[!ht]
 \centering
 \includegraphics[bb=30 70 530 750,width=160mm,clip]{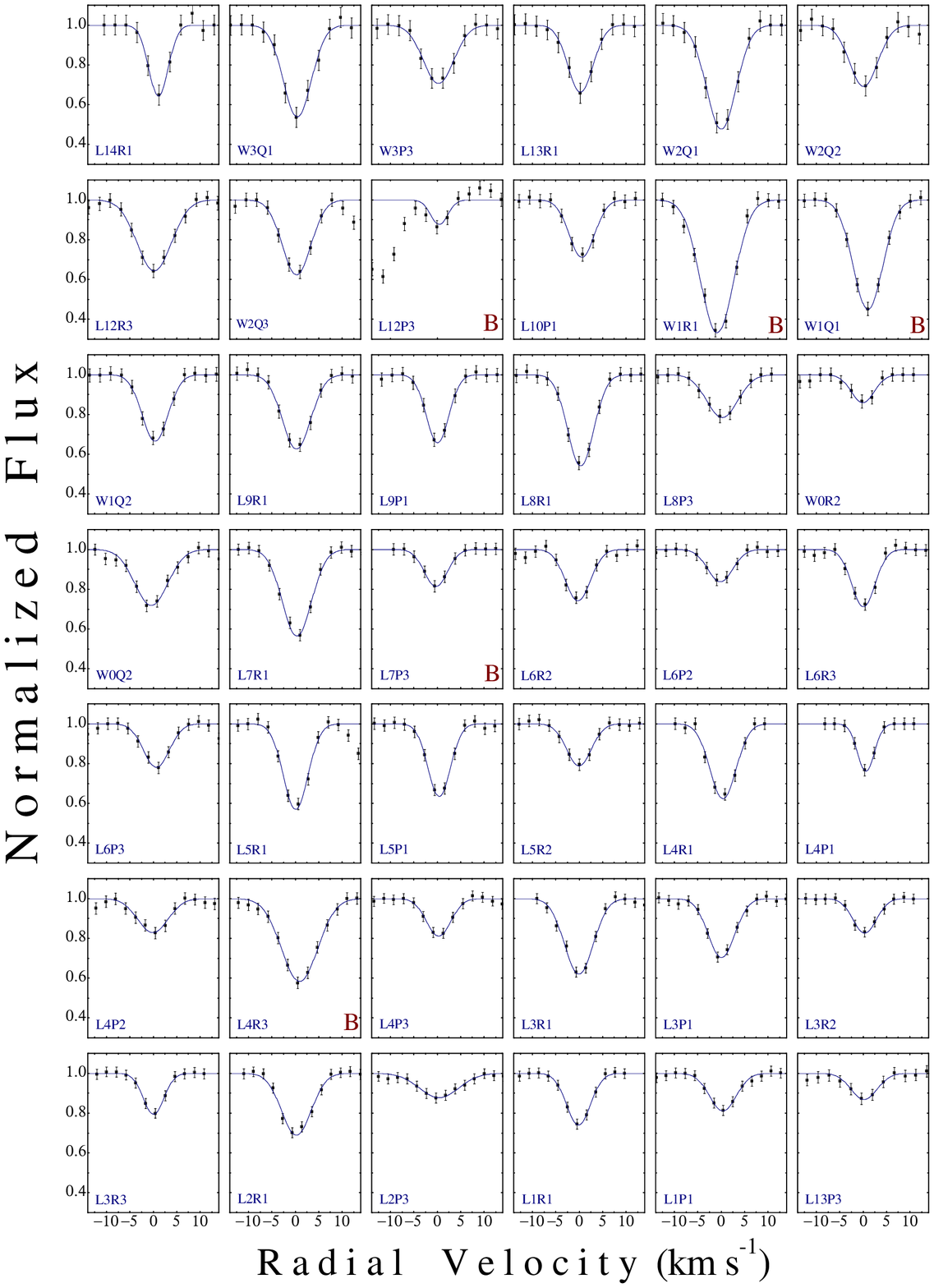}
      \caption{Profiles of selected H$_2$ lines in the absorption system toward Q~0347$-$383.
               The letter ``B'' marks lines which have a good profile but do not satisfy the selection
               criteria (see Section~4, and Fig.~\ref{curv}). The zero point of the radial velocity corresponds
               to the redshift $z_{\rm abs}=3.0249$.}
      \label{prof0347}
   \end{figure*}
\begin{figure*}[!ht]
 \centering
 \includegraphics[bb=30 70 530 750,width=160mm,clip]{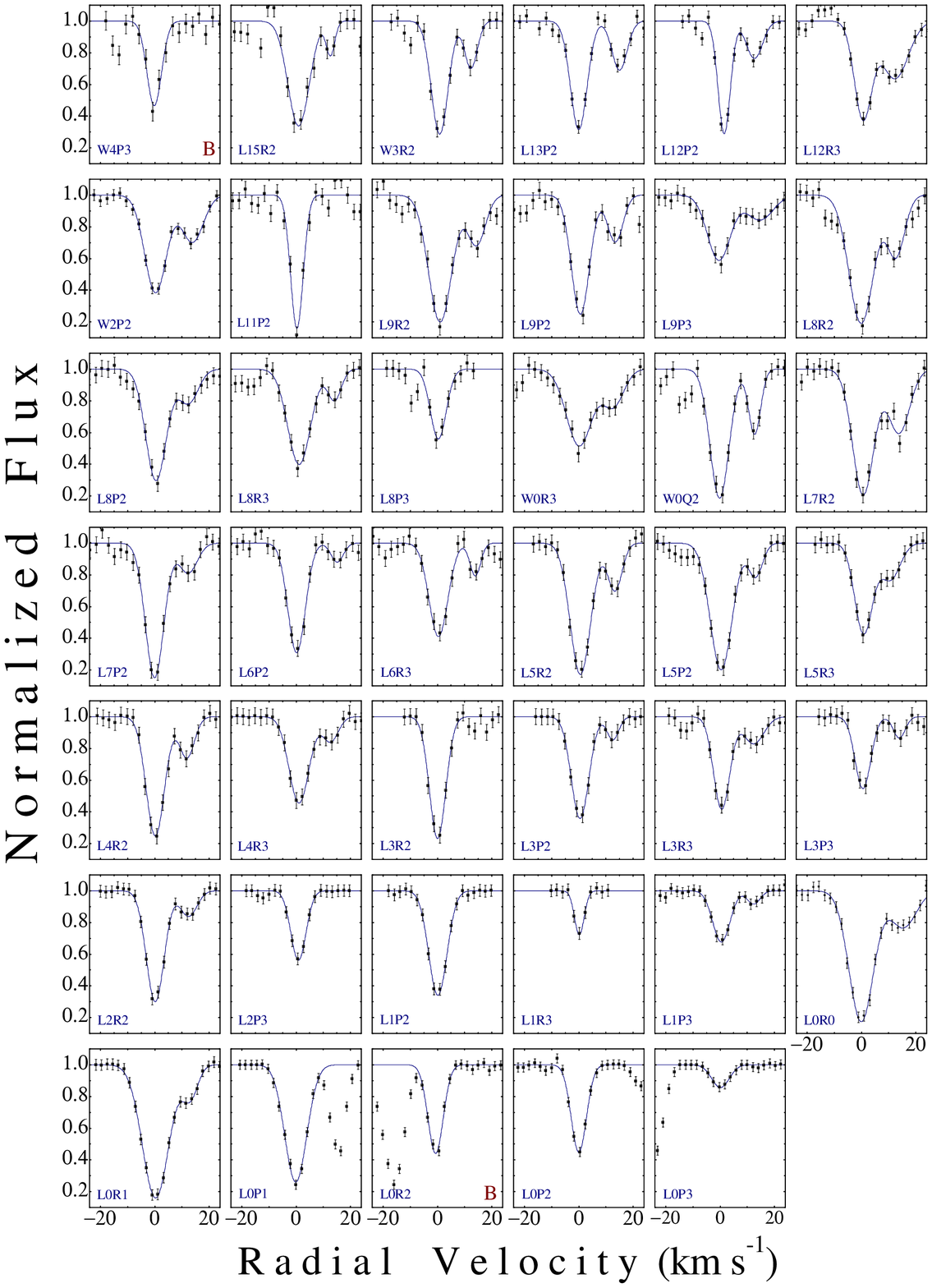}
      \caption{Profiles of selected H$_2$ lines in the absorption system toward Q~0405$-$443.
               The letter ``B'' marks lines which have a good profile but do not satisfy the selection
               criteria (see section~4, and Fig.~\ref{curv}). The zero point of the radial velocity corresponds
               to the redshift $z_{\rm abs}=2.5947$.}
      \label{prof0405}
   \end{figure*}
\begin{figure*}[!ht]
 \centering
 \includegraphics[bb=15 300 555 755,width=16.5cm,clip]{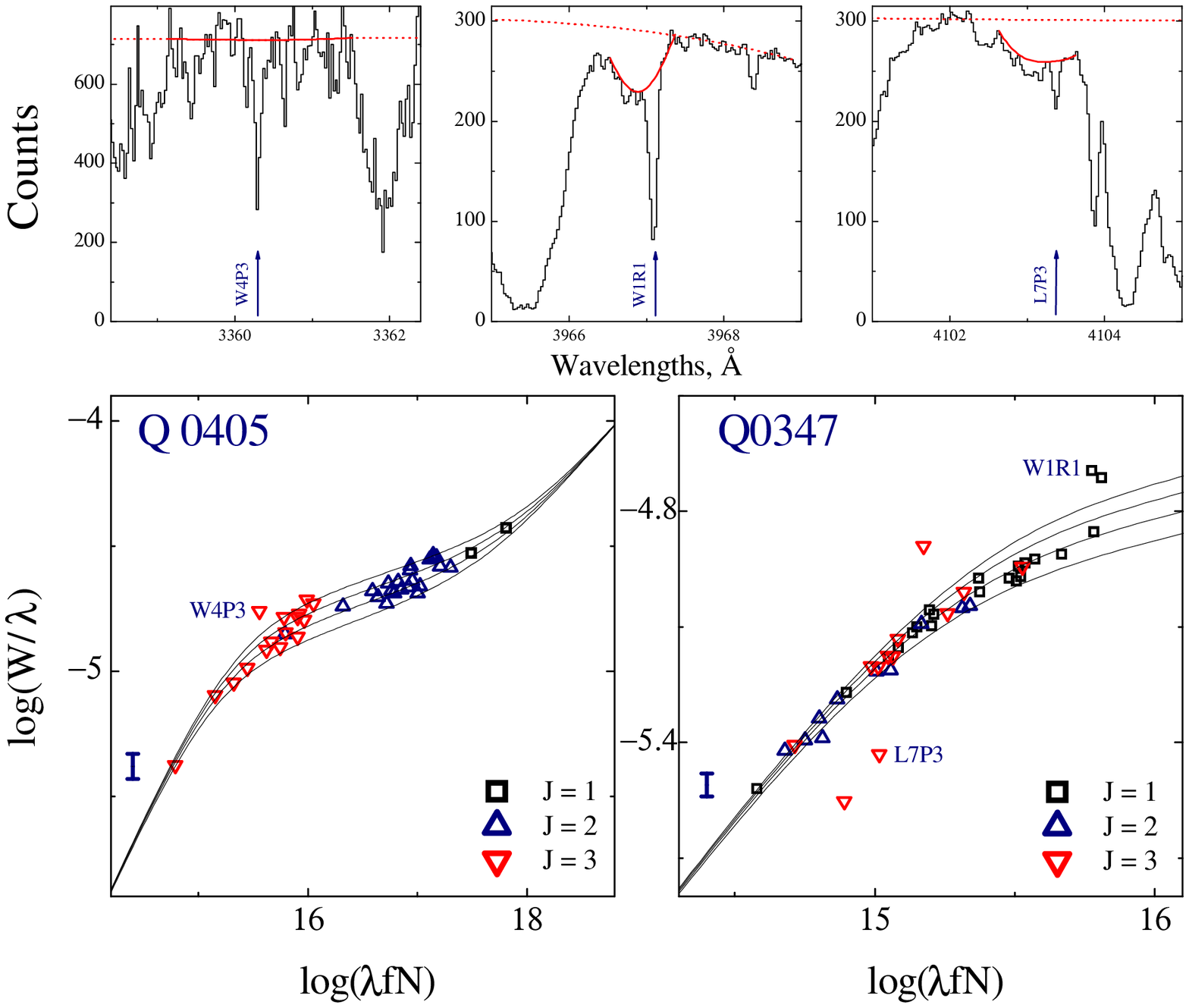}
      \caption{Curves of growth for the observed absorption lines (bottom panels; $\, \lambda$[\AA], N[cm$^{-2}$]).
               The three top panels illustrate
               the procedure we use to fix the local continuum (solid bold line). The equivalent
               widths for the W4P3 and W1R1 lines (first and second top panels) lie above
               the corresponding curves of growth. It means that the lines are probably blended.
               The equivalent widths for L7P3 line (third top panel) lies under the
               curve of growth. It means that the real continuum (dash line) probably lies above the local one.
               In the analysis, we do not use these lines as well as other lines
               that are not on the curve of growth. (The error bars on the bottom panels show
               typical errors for the plotted points).}
         \label{curv}
   \end{figure*}
We used the UVES echelle spectrograph mounted on the Very Large Telescope of
the European Southern Observatory to obtain new and better quality data
(compared to what was available in the UVES data base) on two high-redshift
($z_{\rm em}$~=~3.22 and 3.02) bright quasars, respectively Q~0347$-$383 and
Q~0405$-$443. Nine exposures of 1.5~h each were taken for each of the quasars
over six nights under sub-arcsec seeing conditions in January 2002 and 2003
for, respectively, Q~0347$-$383 and Q~0405$-$443. The slit was 0.8~arcsec wide
resulting in a resolution of R~$\sim$~53000 over the wavelength range
3290-4515~\AA. Thorium-Argon calibration data were taken with different slit
widths (from 0.8 to 1.4~arcsec) before and after each exposure and data
reduction was performed using these different calibration settings to ensure
accurate wavelength calibration. Spectra were extracted using procedures
implemented in MIDAS, the ESO data reduction package. The reduction is
particularly robust as only one CCD is used for the observations (setting
\#390). We have extracted the lamp spectra in the same way as the science
spectra and checked that there is no systematic shift in the position of the
emission lines.

Possible systematic effects leading to wavelength mis-calibration have been
discussed by Murphy et al. (\cite{Murphy}) and we specify here a few technical
points. The wavelength calibration has been extensively checked using ThAr
lamps. Errors measured from the lamp spectra are typically $\sim 2$~m\AA.
Air-vacuum wavelength conversion has been made using the Edl\'en (\cite{edlen})
formula at $15^{\rm o}\,$C. A shift in the wavelength scale can be introduced
if the Thorium-Argon lamp and the science spectra are taken at systematically
different temperatures and pressures. This is not the case here as calibration
spectra were taken just before and after the science exposures. The
temperature variations measured over one night in UVES are smaller than 0.5~K
(see Dekker et al. \cite{dekker}). Heliocentric correction is done using
Stumpff (\cite{Stumpff}) formula. In addition, all exposures were taken
with the slit aligned with the parallactic angle so that atmospheric dispersion
has little effect on our measurements. Therefore, as discussed by Murphy et al.
(\cite{Murphy}), uncertainties due to these effects are neligible.

\section{Data Analysis}
In each of the quasar spectra there is a damped Lyman-$\alpha$
system in which H$_2$ has been well studied, at $z_{\rm
abs}$~=~3.0249 (Levshakov et al. \cite{Lev2}, Ledoux et al.
\cite{led03}), and 2.5947 (Ledoux et al. \cite{led03}) for
Q~0347$-$383 and Q~0405$-$443, respectively. A crucial
advantage of these H$_2$ absorption systems is that numerous
unsaturated lines with narrow simple profiles are seen. A single component
profile is sufficient to fit the lines on the line of sight toward
Q~0347$-$383 and profiles of two well separated
($\Delta V$~=~13~km~s$^{-1}$) components are fitted in the case
of Q~0405$-$443. The absorption lines are
shown in Fig.~\ref{prof0347} and ~\ref{prof0405} for the
two quasars respectively.
For the latter object, we fitted the two components but discuss
only the positions of the strongest one in the following.

\subsection{Selection of lines}
We selected H$_2$ lines that are not obviously blended with other
narrow absorptions, in particular H~{\sc i} intervening
Lyman-$\alpha$ lines, and that have normalized central intensities
larger than 0.1 and smaller than 0.9. The 82 selected lines (42
toward Q~0347$-$383 and 40 toward Q~0405$-$443) were fitted with
Voigt profiles estimating the continuum locally. Possible hidden
inaccuracy or problems related to individual lines (inacurate
continuum determination, possible non-obvious blends, etc...) were
checked by constructing curves of growth for the two sets of
molecular lines (Fig.~\ref{curv}).
For this, we plot log($W/\lambda$) versus log($\lambda f N$) for
the lines observed along both lines of sight (see Fig.~\ref{curv})
together with theoretical curves corresponding to Doppler
parameters $b$~=~1.7, 1.5, 1.3, and 1.1 km~s$^{-1}$. It is
apparent from the figure that 5 lines toward Q~0347$-$383 and one
line toward Q~0405$-$443 do not lie on the theoretical curves.
This is probably a consequence of blending with other lines in the
Lyman-$\alpha$ forest and/or difficulties in positioning the
continuum as illustrated for three of these lines in the top
panels of Fig.~\ref{curv}. The six lines are marked with a sign
"B" in Figs.~1 and 2 and are not considered in the following
analysis which is therefore based on 76 lines (37 toward
Q~0347$-$383 and 39 toward Q~0405$-$443).

For the selected lines, the curves of growth analysis gives the
column density (for each rotational level), $N_{\rm
J}$~[cm$^{-2}$], and the Doppler parameter $b$. For the
Q~0405$-$443 absorption system, $\log N_{\rm J=1} = 17.7\pm0.2$,
$\log N_{\rm J=2} = 15.9\pm0.3$, $\log N_{\rm J=3} = 14.7\pm0.3$,
and $b$~=~1.4$\pm$0.3 km~s$^{-1}$. For the Q~0347$-$383 absorption
system $\log N_{\rm J=1} = 14.3\pm0.2$, $\log N_{\rm J=2} =
13.8\pm0.2$, $\log N_{\rm J=3} = 14.0\pm0.2$, and b~=~1.3$\pm$0.2
km~s$^{-1}$.
\subsection{Line parameters}
The atomic parameters of the selected lines are given in
Tables~\ref{zK1}~and~\ref{zK2}. The first column specifies the
lines. The second one gives the sensitivity coefficients $K_{\rm
i}$. The 3rd and 4th columns present the observed wavelengths and
their errors, $\lambda_{\rm i}$ and $\sigma_{\lambda_i}$. The 5th
and 6th columns give the rest wavelengths, $\lambda_{\rm i}^{\rm
lab}$, as estimated in different laboratory experiments. The first
estimate (called $\lambda^{\rm a}_{\rm i}$) is obtained using
level energies (Jennings et al.~\cite{Jen84},
Dabrowski~\cite{dab84}, Abgrall et al.~\cite{Abgr}, Roncin \&
Launay~\cite{Ron}) determined from laboratory observations of the
H$_2$ emission spectrum. Uncertainties for the ground state
energies are less than $3\times10^{-4}$~cm$^{-1}$ (Jennings et
al.~\cite{Jen84}). Uncertainties are more difficult to estimate
for the upper levels although close to 0.1~cm$^{-1}$
(Dabrowski~\cite{dab84}) for most of the lines. This means that
most of the errors are of the same order of magnitude as our
observational measurements, 1~m\AA~ in the rest frame (or about 3
to 4 m\AA~ for the observer). The second estimate (called
$\lambda^{\rm p}_{\rm i}$) is a direct measurement using a
narrow-band tunable extreme UV laser source (Philip et al. 2004).
This experiment is supposed to be much more precise and errors
should be less than 0.011~cm$^{-1}$.
\par\noindent
Observational errors ($\sigma_{\lambda_i}$) are of the order of
3~m\AA, they characterize only the accuracy of the profile fitting
of the observed lines by gaussian profiles. The total error of the
line centrum position can be estimate from the real dispersion
of points (e.g. Fig.5.)
$\sigma_{\lambda_i} \la \lambda^{lab} \sigma_{z_i} \approx 5$m\AA.

\subsection{Consistency of the two lines of sight}

\begin{figure}[!h]
 \centering
 \includegraphics[bb=20 395 550 755,width=90mm,clip]{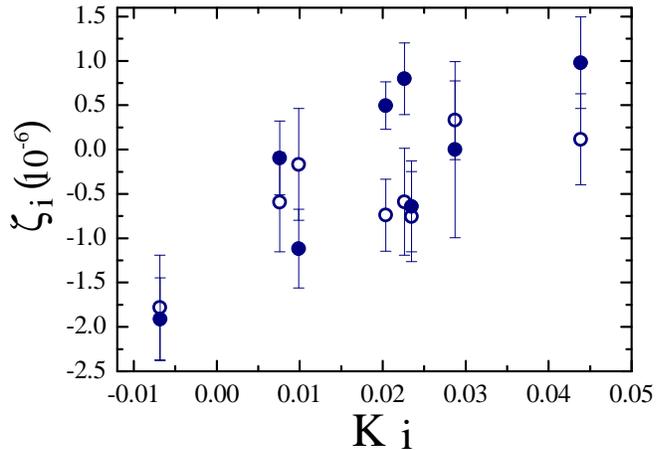}
      \caption{The relative positions, $\zeta_{\rm i}=(z_{\rm i}-\bar{z})/(1+\bar{z})$, of
      lines observed in the spectra of both quasars Q~0405$-$443 (filled circles)
      and Q~0347$-$383 (open circles) are plotted versus the sensitivity coefficient $K_{\rm i}$.
      Here, $\bar{z}$~is the median redshift of all H$_2$ lines observed in the one
      corresponding spectrum. The lines are L3R3, L3R2, L5R2, L6R3, L6P2, W0Q2,
      L8P3, and L12R3 and $\lambda_{\rm lab}^{\rm a}$ are used.
      The 1~$\sigma$ error bars shown are observational (without laboratory errors).
      They are of the order of 3~m\AA~ (in the observer's frame). It can be seen that
      both sets of measurements are mutually consistent within observational errors.
      The fact that the two independent observational measurements agree well indicates
      that our data calibration and measurement procedure are reliable at the level
      required for the study.}
 \label{commL}
\end{figure}

An important internal check of the data quality consists in
comparing measurements of the eight lines present in both QSO
spectra. This is done in Fig.~\ref{commL} where the relative
positions, $\zeta_{\rm i}=(z_{\rm i}^{\rm
obs}-\bar{z})/(1+\bar{z})\;$, is plotted versus $K_{\rm i}$;
$\bar{z}$ being the median redshift (i.e. model independent) of
all H$_2$ lines observed in one spectrum. It can be seen that all
measurements are within observational errors (at the 2$\sigma$ level).
As the two lines of sight have been observed and reduced independently,
this shows that the data calibration and the measurement procedure are
reliable at the level required for the study.
%

\begin{table}[!t]
      \caption[]{~Parameters of H$_2$ lines for the absorption system at~z=3.02490 in the spectrum of Q~0347-383.}
         \label{zK1}
\vspace{-5mm}
     $$
{\scriptsize \tabcolsep=1.7mm
            \begin{tabular}{rrccrr}
            \hline \hline
            \noalign{\smallskip}
            Lines~ & $K_i$~~~ & $\lambda_i$, \AA & $\sigma_{\lambda_i}$, \AA
            & $\lambda^{a}_i$, \AA~~ & $\lambda^{p}_i$, \AA~~ \\
            \noalign{\smallskip}
            \hline
            \noalign{\smallskip}

 L14R1 &  0.05495 & 3811.506 & 0.0025 &  946.9795 &  946.98040\\
 W~3Q1 &  0.02176 & 3813.279 & 0.0022 &  947.4212 &  947.42188\\
 W~3P3 &  0.01724 & 3830.382 & 0.0023 &  951.6711 &  951.67186\\
 L13R1 &  0.05109 & 3844.044 & 0.0024 &  955.0649 &  955.06582\\
 L13P3 &  0.04574 & 3865.717 & 0.0027 &  960.4497 &  960.45063\\
 W~2Q1 &  0.01423 & 3888.435 & 0.0013 &  966.0951 &  966.09608\\
 W~2Q2 &  0.01301 & 3893.205 & 0.0020 &  967.2800 &  967.28110\\
 L12R3 &  0.04386 & 3894.800 & 0.0020 &  967.6758 &  967.67695\\
 W~2Q3 &  0.01120 & 3900.325 & 0.0015 &  969.0481 &  969.04922\\
 L10P1 &  0.04053 & 3955.815 & 0.0017 &  982.8339 &  982.83533\\
 W~1Q2 &  0.00394 & 3976.499 & 0.0035 &  987.9743 &  987.97445\\
 L~9R1 &  0.03796 & 3992.767 & 0.0017 &  992.0156 &  992.01637\\
 L~9P1 &  0.03719 & 3995.960 & 0.0010 &  992.8093 &  992.80968\\
 L~8R1 &  0.03444 & 4034.769 & 0.0008 & 1002.4509 & 1002.45210\\
 L~8P3 &  0.02872 & 4058.651 & 0.0018 & 1008.3849 & 1008.38615\\
 W~0R2 & -0.00503 & 4061.220 & 0.0030 & 1009.0244 & 1009.02492\\
 W~0Q2 & -0.00686 & 4068.922 & 0.0024 & 1010.9389 & 1010.93844\\
 L~7R1 &  0.03062 & 4078.983 & 0.0017 & 1013.4364 & 1013.43701\\
 L~6R2 &  0.02496 & 4131.669 & 0.0028 & 1026.5281 & --\\
 L~6P2 &  0.02347 & 4138.020 & 0.0021 & 1028.1055 & 1028.10609\\
 L~6R3 &  0.02262 & 4141.563 & 0.0025 & 1028.9856 & --\\
 L~6P3 &  0.02053 & 4150.451 & 0.0014 & 1031.1917 & 1031.19260\\
 L~5R1 &  0.02183 & 4174.421 & 0.0020 & 1037.1490 & 1037.14992\\
 L~5P1 &  0.02088 & 4178.483 & 0.0015 & 1038.1568 & 1038.15713\\
 L~5R2 &  0.02038 & 4180.622 & 0.0017 & 1038.6901 & 1038.69027\\
 L~4R1 &  0.01681 & 4225.987 & 0.0026 & 1049.9592 & 1049.95976\\
 L~4P1 &  0.01580 & 4230.303 & 0.0025 & 1051.0317 & 1051.03253\\
 L~4P2 &  0.01369 & 4239.360 & 0.0040 & 1053.2841 & 1053.28426\\
 L~4P3 &  0.01071 & 4252.193 & 0.0010 & 1056.4709 & 1056.47143\\
 L~3R1 &  0.01132 & 4280.316 & 0.0020 & 1063.4594 & 1063.46014\\
 L~3P1 &  0.01026 & 4284.928 & 0.0015 & 1064.6048 & 1064.60539\\
 L~3R2 &  0.00989 & 4286.499 & 0.0027 & 1064.9950 & 1064.99481\\
 L~3R3 &  0.00758 & 4296.491 & 0.0024 & 1067.4780 & 1067.47855\\
 L~2R1 &  0.00535 & 4337.628 & 0.0025 & 1077.6979 & 1077.69894\\
 L~2P3 & -0.00098 & 4365.242 & 0.0027 & 1084.5593 & 1084.56034\\
 L~1R1 & -0.00113 & 4398.132 & 0.0015 & 1092.7316 & --\\
 L~1P1 & -0.00234 & 4403.449 & 0.0030 & 1094.0516 & --\\
       &          &          &        &           & \\
       &          &          &        &           & \\
            \noalign{\smallskip}
            \hline
         \end{tabular} }
     $$

$^a$ H$_2$ laboratory wavelengths are from (Jennings et
al.~\cite{Jen84}, Dabrowski~\cite{dab84}, Abgrall et
al.~\cite{Abgr}, Roncin \& Launay~\cite{Ron}). \\
$^p$ H$_2$ laboratory wavelengths are from (Philip et al. 2004).
   \end{table}
%

\begin{table}[!t]
      \caption[]{~Parameters of H$_2$ lines for the absorption system at~z=2.59473 in the spectrum of Q~0405-443.}
         \label{zK2}
\vspace{-5mm}
     $$
{\scriptsize \tabcolsep=1.7mm
            \begin{tabular}{rrccrr}
            \hline \hline
            \noalign{\smallskip}
            Lines~ & $K_i$~~~ & $\lambda_i$, \AA & $\sigma_{\lambda_i}$, \AA
            & $\lambda^{a}_i$, \AA~~ & $\lambda^{p}_i$, \AA~~ \\
            \noalign{\smallskip}
            \hline
            \noalign{\smallskip}
L15R2 & 0.05816 & 3381.302 & 0.0030 &  940.6257 &  --\\
W~3R2 & 0.02193 & 3404.619 & 0.0019 &  947.1116 &  947.11169\\
L13P2 & 0.04848 & 3442.500 & 0.0025 &  957.6516 &  957.65223\\
L12P2 & 0.04503 & 3473.513 & 0.0023 &  966.2751 &  966.27550\\
L12R3 & 0.04386 & 3478.541 & 0.0018 &  967.6758 &  967.67695\\
W~2P2 & 0.01198 & 3480.759 & 0.0015 &  968.2943 &  968.29522\\
L11P2 & 0.04177 & 3506.107 & 0.0023 &  975.3454 &  975.34576\\
L~9R2 & 0.03647 & 3571.558 & 0.0018 &  993.5505 &  993.55061\\
L~9P2 & 0.03519 & 3576.309 & 0.0022 &  994.8735 &  994.87408\\
L~9P3 & 0.03232 & 3586.919 & 0.0035 &  997.8264 &  997.82717\\
L~8R2 & 0.03296 & 3609.057 & 0.0034 & 1003.9851 & 1003.98545\\
L~8P2 & 0.03161 & 3614.121 & 0.0025 & 1005.3923 & 1005.39320\\
L~8R3 & 0.03061 & 3617.795 & 0.0018 & 1006.4130 & 1006.41416\\
L~8P3 & 0.02872 & 3624.876 & 0.0036 & 1008.3849 & 1008.38615\\
W~0R3 &-0.00617 & 3631.147 & 0.0025 & 1010.1304 & 1010.13025\\
W~0Q2 &-0.00682 & 3634.050 & 0.0017 & 1010.9389 & 1010.93844\\
L~7R2 & 0.02914 & 3648.575 & 0.0025 & 1014.9767 & 1014.97685\\
L~7P2 & 0.02772 & 3653.904 & 0.0024 & 1016.4612 & 1016.46125\\
L~6P2 & 0.02347 & 3695.764 & 0.0019 & 1028.1056 & 1028.10609\\
L~6R3 & 0.02262 & 3698.933 & 0.0015 & 1028.9856 & --\\
L~5R2 & 0.02038 & 3733.817 & 0.0010 & 1038.6901 & 1038.69027\\
L~5P2 & 0.01880 & 3739.844 & 0.0010 & 1040.3672 & 1040.36732\\
L~5R3 & 0.01805 & 3742.691 & 0.0017 & 1041.1583 & 1041.15892\\
L~4R2 & 0.01536 & 3779.856 & 0.0021 & 1051.4988 & 1051.49857\\
L~4R3 & 0.01304 & 3788.770 & 0.0015 & 1053.9753 & 1053.97610\\
L~3R2 & 0.00989 & 3828.370 & 0.0017 & 1064.9950 & 1064.99481\\
L~3P2 & 0.00812 & 3835.227 & 0.0020 & 1066.9006 & 1066.90068\\
L~3R3 & 0.00758 & 3837.300 & 0.0016 & 1067.4781 & 1067.47855\\
L~3P3 & 0.00511 & 3846.871 & 0.0023 & 1070.1401 & 1070.14087\\
L~2R2 & 0.00394 & 3879.528 & 0.0021 & 1079.2259 & 1079.22491\\
L~2P3 &-0.00098 & 3898.707 & 0.0024 & 1084.5593 & 1084.56034\\
L~1P2 &-0.00453 & 3941.405 & 0.0020 & 1096.4390 & --\\
L~1R3 &-0.00479 & 3942.435 & 0.0029 & 1096.7246 & --\\
L~1P3 &-0.00760 & 3953.441 & 0.0014 & 1099.7864 & --\\
L~0R0 &-0.00772 & 3983.420 & 0.0020 & 1108.1277 & --\\
L~0R1 &-0.00818 & 3985.240 & 0.0018 & 1108.6328 & --\\
L~0P1 &-0.00948 & 3990.370 & 0.0020 & 1110.0617 & --\\
L~0P2 &-0.01170 & 3999.123 & 0.0013 & 1112.4963 & --\\
L~0R3 &-0.01178 & 3999.437 & 0.0020 & 1112.5830 & --\\
            \noalign{\smallskip}
            \hline
         \end{tabular} }
     $$
$^a$ H$_2$ laboratory wavelengths are from (Jennings et
al.~\cite{Jen84}, Dabrowski~\cite{dab84}, Abgrall et
al.~\cite{Abgr}, Roncin \& Launay~\cite{Ron}). \\
$^p$ H$_2$ laboratory wavelengths are from (Philip et al. 2004).
   \end{table}

\section{Results}

\begin{figure*}[!ht]
 \centering
 \includegraphics[bb=15 180 565 755,width=180mm,clip]{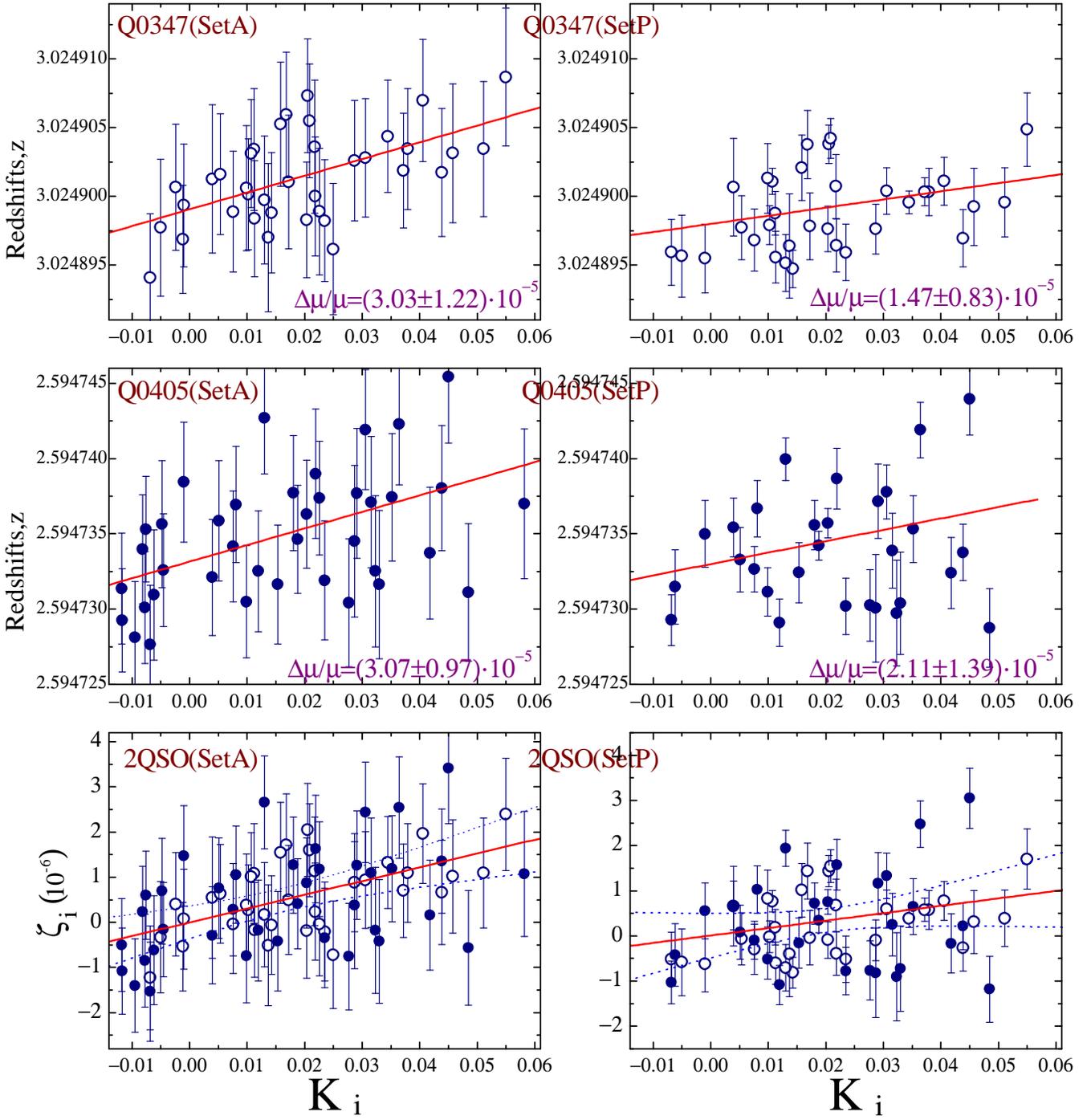}
     \caption{Regression analysis using rest wavelengths from
energy levels (Sample A; left column) and from laser experiments
(Sample P; right column) for both quasars (top and middle rows)
and the whole sample (bottom row).}
      \label{mu3}
   \end{figure*}

In Fig.~\ref{mu3} we plot $z_{\rm i}$ versus $K_{\rm i}$ for
absorption lines observed in the spectra of Q~0347$-$383 (open
circles) and Q~0405$-$443 (filled circles) respectively. The left
hand side panels corresponds to rest-wavelengths $\lambda_{\rm
lab}^{\rm a}$ and the right hand side panel to rest-wavelengths
$\lambda_{\rm lab}^{\rm p}$. The best fit of the linear regression
$z_{\rm i}$-to-$K_{\rm i}$ in accordance with Eq.~(\ref{korr}) is
overplotted in all panels.
\par\noindent
For the left hand side panels, error bars are the combination of
measurement errors (evaluated from col.~5 of
Tables~\ref{zK1}~and~\ref{zK2}) with an error of 1~m\AA, in the
rest frame, to account for uncertainties in laboratory wavelength
determination. For the right hand side panels, uncertainties in
laboratory wavelength determination are supposed to be of the
order of 0.1~m\AA.
\par\noindent
Data for both quasars are combined in the bottom panels using the
following formula for reduced redshift $\zeta_{\rm i}$:
\begin{equation}
\zeta_{\rm i}=\frac{z_{\rm i}^{\rm obs}-z_{\rm abs}}{1+z_{\rm
abs}}
\end{equation}
In this formula, $z_{\rm abs}$ is obtained from the best linear
fit of the data in accordance with Eq.~(\ref{korr}). They are
$z_{\rm abs}=3.02489904(120)$ for Q~0347$-$383 and $z_{\rm
abs}=2.59473315(81)$ for Q~0405$-$443.

The results of the linear regression analysis are presented in
Table~\ref{R_mu}. The first column gives the QSO name, the second
one gives the number of lines used in the regression analysis, the
third column gives the estimated value of $\Delta \mu/\mu$.
Estimates of $\Delta \mu/\mu$ are given using the two sets of rest
wavelengths for both quasars separately as well as for the
combined sample. Most of the lines are from J~=~1 for Q~0347$-$383
and J~=~2 for Q~0405$-$443 and the results of the regression
analysis for these two subsamples are also given. Note that the
number of lines is smaller in the case where $\lambda_{\rm
lab}^{\rm p}$ are used because not all the lines have been
measured.

   \begin{table}[!ht]
      \caption[]{$\Delta \mu / \mu$ estimates from different samples.}
         \label{R_mu}
\vspace{-5mm}
     $$
         \begin{array}{lcc}
            \hline \hline
            \noalign{\smallskip}
                & ~\mbox{N.~of~L.}~ & ~\Delta\mu/\mu~ \\
            \noalign{\smallskip}
            \hline \noalign{\smallskip}
               & \mbox{Set (A)} &  \\
            \hline
            \noalign{\smallskip}

 ~\mbox{Q~0347$-$383}~ & ~~37~~ & ~(3.03\pm1.22)\times 10^{-5}~ \\

            \noalign{\smallskip}
            \noalign{\smallskip}

 ~\mbox{Q~0347$-$383~~(J=1)}~ & ~~18~~ & ~(3.23\pm1.63)\times 10^{-5}~ \\

            \noalign{\smallskip}
            \noalign{\smallskip}

 ~\mbox{Q~0405$-$443}~ & ~~39~~ & ~(3.07\pm0.97)\times 10^{-5}~ \\

            \noalign{\smallskip}
            \noalign{\smallskip}

 ~\mbox{Q~0405$-$443~~(J=2)}~ & ~~22~~ & ~(3.78\pm1.36)\times 10^{-5}~ \\

            \noalign{\smallskip}
            \noalign{\smallskip}
 ~\mbox{2~QSO}~     & ~~76~~ & ~(3.05\pm0.75)\times 10^{-5}~ \\
            \noalign{\smallskip}

            \hline \noalign{\smallskip}
               & \mbox{Set (P)} &  \\
            \hline
            \noalign{\smallskip}

~\mbox{Q~0347$-$383}~ & ~~33~~ & ~(1.47\pm0.83)\times 10^{-5}~ \\

            \noalign{\smallskip}
            \noalign{\smallskip}

~\mbox{Q~0405$-$443}~ & ~~29~~ & ~(2.11\pm1.39)\times 10^{-5}~ \\

            \noalign{\smallskip}
            \noalign{\smallskip}

 ~\mbox{2~QSO}~     & ~~62~~ & ~(1.65\pm0.74)\times 10^{-5}~ \\
            \noalign{\smallskip}
            \hline\hline

         \end{array}
     $$
   \end{table}

Systematic effects in measurements of the central position of a
line profile were discussed by Ivanchik et al. (\cite{IAV2}) and
in more detail by Murphy et al. (\cite{Murphy}). Here we discuss
two possible sources of systematic errors more specifically.


\begin{figure*}[!ht]
 \centering
 \includegraphics[bb=25 425 555 755,width=160mm,clip]{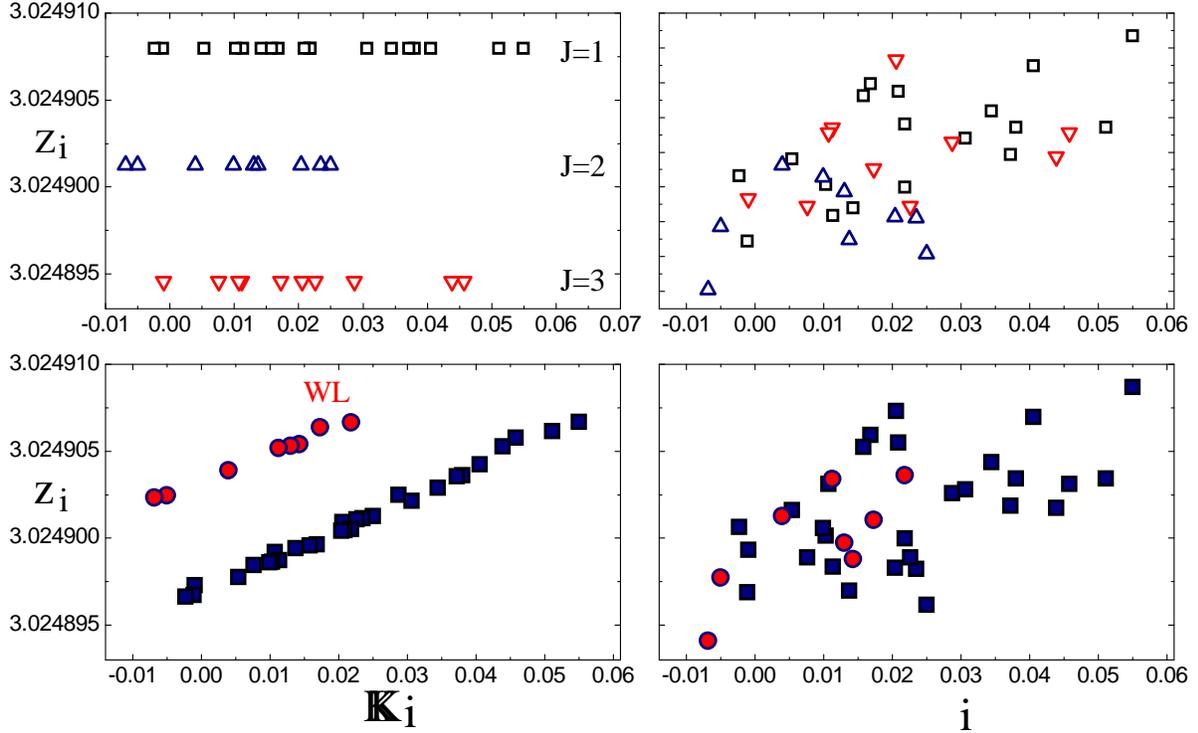}
      \caption{Possible systematic effects. The left top panel illustrates
               the kinematic effect that could arise if absorption lines from
               different rotational levels J had different velocity positions in the spectrum
               (Jenkins \& Peimbert, \cite{JP}). It may mimic
               $\mu$-variation if ranges of $K_{\rm i}$ for different J levels do not overlap.
               The observed situation is shown in the right top panel.
               The $K_{\rm i}$ ranges for J~=~1 and J~=~3 overlap enough
               and slopes are similar for each J levels within the statistical uncertainties.
               The second type of systematic error could be a consequence of any effect that
               could produce a shift increasing monotonously with wavelength:
               H$_2$ laboratory wavelengths, and/or Th-Ar calibration,
               air-vacuum wavelength conversion, atmospheric dispersion effects,
               instrumental profile variation with $\lambda$ etc... (Murphy et al., \cite{Murphy})).
               In that case, the regression lines for Lyman and Werner bands should be shifted
               from one to the other (bottom left panel). It is apparent from the observation (bottom
               right panel) that such an effect cannot be the dominant source of the correlation.}

         \label{Shift}
\end{figure*}

The first one may be called the kinematic effect. Due to peculiar
structure in the clouds H$_2$ molecular features from different
rotational levels J~=~0, 1, 2, 3... may not be produced in the
same region of the absorbing cloud and therefore may have
different mean observed velocities (e.g. Jenkins \& Peimbert,
\cite{JP}). This could lead to relative shifts between the common
redshifts derived for lines from different rotational levels J.
This is illustrated in Fig.~\ref{Shift}. The top left panel shows
the ideal $z$-$K_{\rm i}$ relation (for $\Delta\mu$~=~0) for a
sample of J~=~1, 2, 3 lines corresponding to the lines observed
toward Q~0347$-$383 for which we have imposed a shift of
0.5~km~s$^{-1}$ between different J levels. This effect could
mimic $\mu$-variation if ranges of $K_{\rm i}$ for different J
levels do not overlap. The observed situation is shown in the
top-right panel of the same figure. In that case, the overlap
between $K_{\rm i}$ ranges for J~=~1 and J~=~3 lines is important
enough so that the correlation cannot be due to this effect.
Moreover, most of the lines (18 out of 37) are from the J~=~1
level and the linear regression analysis for these lines only
gives a $\Delta\mu/\mu$ value similar to what is derived
from the whole sample (see Tables~\ref{R_mu}).

Another systematic error could be produced by any effect producing
a shift monotonically increasing with increasing wavelength. This
could be a consequence of slightly unprecise Th-Ar calibration or
air-vacuum wavelength conversion, or atmospheric dispersion
effects, instrumental profile variation etc... (Murphy et al.,
\cite{Murphy}). Indeed, there is a well-known correlation between
$K_{\rm i}$ and $\lambda_0$. Such effects could lead to a slope in
the regression line, i.e. mimic $\mu$-variation. This is
illustrated in the bottom-left panel of Fig.~\ref{Shift} where
such an ideal artificial effect has been applied to the sample of
lines seen toward Q~0347$-$383. It can be seen however that the
Werner and Lyman-band lines have different locations in the plane.
The reason is that for the same $K_{\rm i}$ coefficient, the
Werner lines have larger $\lambda_0$. It is apparent from the
observed sample (bottom-right panel) that there is no such shift.

\section{Conclusion}

Using 76 H$_2$ absorption lines observed at $z_{\rm
abs}$~=~2.59473 and 3.02490 in the spectra of two quasars,
respectively, Q~0405$-$443 and Q~0347$-$383, we have searched for
any correlation between the relative positions of H$_2$ absorption
lines measured as $\zeta_{\rm i}$~=~($z_{\rm i}-{\bar
z}$)/(1+${\bar z}$) and the sensitivity coefficients $K_{\rm i}$
of the lines to a change in $\mu$. A positive correlation could be
interpreted as a variation of the proton-to-electron mass ratio,
$\Delta\mu$/$\mu$. We use two sets of rest wavelengths as
estimated from different laboratory experiments. Wavelengths
derived from energy level determination give
$\Delta\mu$/$\mu$~=~(3.05$\pm$0.75)$\times$10$^{-5}$, over the
past $\sim$12 Gyrs. However, wavelengths derived from a direct
and, in principle, more precise determination using laser techniques
give $\Delta\mu$/$\mu$~=~(1.64$\pm$0.74)$\times$10$^{-5}$. The
latter limit is the most stringent limit obtained to date now on the
variation of this fundamental constant. This limit, together with
the limit on $\Delta \alpha/\alpha$, yields
an estimate of the $R$ parameter defined as $\Delta\mu/\mu\simeq R
\Delta\alpha/\alpha$. Using $\Delta\alpha/\alpha$ from Murphy et
al. (\cite{Murphy03}) gives $\,-9.5\leq R \leq-0.2\,$, and $\Delta
\alpha/\alpha$ from Chand et al. (\cite{Chand}) gives $|R|>1$ (at
the $2\sigma$ C.L.).


\begin{acknowledgements}
We thank F. Roncin, H. Abgrall, E. Roueff for useful discussions,
and S.A. Levshakov for useful remarks. PPJ and RS gratefully
acknowledge support from the Indo-French Centre for the Promotion
of Advanced Research (Centre Franco-Indien pour la Promotion de la
Recherche Avanc\'ee) under contract No. 3004-A. AI and DV are
grateful for the support by the RFBR grant (03-07-90200) and
the grant of Leading scientific schools (NSh-1115.2003.2).
\end{acknowledgements}


\begin{thebibliography}{}

\bibitem[1993]{Abgr} Abgrall,~H., Roueff,~E., Launay,~F., Roncin,~J.-Y.,
\& Subtil,~J.-L. 1993, J. of Mol. Spec., 157, 512

\bibitem[2002]{Calmet}
Calmet,~X., \& Fritzsch,~H. 2002, Eur. Phys. J., C24, 639

\bibitem[2004]{Chand}
Chand,~H., Srianand,~R., Petitjean,~P., \& Aracil,~B. 2004, A\&A,
417, 853

\bibitem[1995]{Cowie}
Cowie,~L., \& Songaila,~A. 1995, ApJ, 453, 596

\bibitem[1984]{dab84} Dabrowski,~I. 1984, Can. J. Phys., 62, 1639

\bibitem[2000]{dekker} Dekker, H., D'Odorico, S., Kaufer, A., Delabre, B., \&
Kotzlowski, H. 2000, SPIE 4008, 534

\bibitem[2003]{Dent} Dent,~T., \& Fairbairn,~M. 2003, Nuc. Phys. B, 653, 256

\bibitem[2003]{Dine} Dine,~M., Nir,~Y., Raz,~G., \& Volansky,~T. 2003, Phys.
Rev. D, 67, 015009

\bibitem[1999]{Dzuba}
Dzuba,~V., Flambaum,~V., \& Webb,~J. 1999, Phys. Rev. A., 59, 230

\bibitem[1966]{edlen} Edl\'en,~B. 1966, Metrologia, 2, 71

\bibitem[1999]{IAV2}
Ivanchik, A., Potekhin,~A., \& Varshalovich, D. 1999, A\&A, 343,
439

\bibitem[2003]{IAV}
Ivanchik, A., Petitjean, P., Rodriguez, E., \& Varshalovich, D.
2003, Astrophys. Space Sci., 283, 583

\bibitem[1997]{JP}
Jenkins,~E.B., \& Peimbert,~A. 1997, ApJ, 477, 265

\bibitem[1984]{Jen84} Jennings,~D.E., Bragg,~S.L., \& Brault,~J.W. 1984,
ApJ, 282, L85

\bibitem[2002]{Langacker}
Langacker,~P., Segre,~G., \& Strassler,~M. 2002, Phys. Lett.,
B528, 121

\bibitem[2003]{led03} Ledoux, C., Petitjean, P., \& Srianand, R., 2003,
MNRAS, 346, 209

\bibitem[2002a]{Lev1}
Levshakov, S., Dessauges-Zavadsky, M., D'Odorico, S., \& Molaro,
P. 2002a, MNRAS, 333, 373

\bibitem[2002b]{Lev2}
Levshakov, S., Dessauges-Zavadsky, M., D'Odorico, S., \& Molaro,
P. 2002b, ApJ, 565, 696

\bibitem[2000]{Mohr} Mohr, P.J., \& Taylor, B.N. 2002,
CODATA Recommended Values of the Fundamental Physical Constants:
2002, (http://physics.nist.gov/), to be published.

\bibitem[2001]{Murphy}
Murphy, M., Webb,~J., Flambaum,~V., Churchill,~C., \&
Prochaska,~J. 2001, MNRAS, 327, 1223

\bibitem[2003]{Murphy03}
Murphy, M., Webb,~J., \& Flambaum,~V. 2003, MNRAS, 345, 609

\bibitem[2002]{Olive}
Olive,~K.A., Pospelov,~M., Qian,~Y.-Z., Coc,~A., Casse,~M., \&
Vangioni-Flam,~E. 2002, Phys. Rev., D66, 045022

\bibitem[2004]{Philip}
Philip,~J., Sprengers,~J.P., Pielage,~Th., Lange,~C.A.,
Ubachs,~W., \& Reinhold,~E. 2004, Can. J. Chem., 82, 713

\bibitem[1998]{PIV}
Potekhin, A., Ivanchik, A., Varshalovich, D., Lanzetta, K.,
Baldwin, J., Williger, G., \& Carswell, R. 1998, ApJ, 505, 523

\bibitem[2004]{Quast} Quast,~R., Reimers,~D., \& Levshakov, S. 2004,
A\&A, 415, L7

\bibitem[1994]{Ron}
Roncin, J.-Y. \& Launay, F. 1994, Journal Phys. and Chem.
Reference Data, No.~4

\bibitem[2004]{Srianand}
Srianand,~R., Chand,~H., Petitjean,~P., \& Aracil,~B. 2004, PRL,
92, 121302

\bibitem[1980]{Stumpff} Stumpff,~P. 1980, A\&AS, 41, 1

\bibitem[2004]{Ubachs} Ubachs,~W., \& Reinhold,~E. 2004, PRL,
92, 101302

\bibitem[1993]{VarshLev} Varshalovich, D., \& Levshakov, S. 1993,
JETP Letters, 58, 231

\bibitem[1995]{VarshPo} Varshalovich, D., \& Potekhin, A. 1995,
Space Sci. Rev., 74, 259

\bibitem[1999]{Webb1}
Webb,~J., Flambaum,~V., Churchill,~C., Drinkwater,~M., \&
Barrow,~J. 1999, Phys. Rev. Lett., 82, 884

\end{thebibliography}
\end{document}